\documentclass[11pt,a4paper]{jpconf}
\usepackage{graphicx}
\def\MSbar{\relax\ifmmode\overline                        
            {\rm MS}\else{$\overline{\rm MS}${ }}\fi}     
\begin{document}
\title{The generalized BLM approach to fix scale-dependence in QCD:\\
the current status of investigations}

\author{A L Kataev$~1$}

\address{$~1$
Institute for Nuclear Research of the Academy of Sciences of
Russia,  Moscow 117312,  Russia}

\ead{kataev@ms2.inr.ac.ru}

\begin{abstract}
I present a brief review  of the
generalized Brodsky-Lepage-McKenzie (BLM)  approaches to
fix the  scale-dependence of 
the renormalization group (RG)  invariant  quantities in QCD. 
At first, these approaches are  based on the
expansions of the coefficients of the perturbative series
for the RG-invariant quantities in the products of the  coefficients $\beta_i$
of  the  QCD $\beta$-function, 
which are evaluated in the MS-like schemes. 
As a next step all $\beta_i$-dependent terms
are absorbed into the BLM-type scale(s) of the powers of the QCD couplings.
The difference between two existing formulations of the
above mentioned generalizations 
based on the seBLM approach and the Principle of Maximal Conformality (PMC) 
are clarified
in the case of  the Bjorken polarized deep-inelastic scattering  sum rule.
Using the conformal symmetry-based  relations
for the non-singlet  coefficient functions of the
 Adler D-function and of  Bjorken polarized
deep-inelastic scattering sum rules
$C^{\rm Bjp}_{\rm NS}(a_s)$
the   $\beta_i$-dependent structure of the NNLO approximation
for $C^{\rm Bjp}_{\rm NS}(a_s)$  is predicted in QCD with $n_{gl}$-multiplet
of  gluino degrees of freedom, which appear in  SUSY extension of QCD.
The importance of performing the analytical   
calculation of  the  N$^3$LO   
additional contributions of   $n_{gl}$ gluino multiplet 
to $C^{\rm Bjp}_{\rm NS}(a_s)$ for checking  the presented in the 
report    NNLO prediction and  
for the  studies of the possibility to determine  the discussed  
$\{\beta\}$-expansion  pattern of this sum rule at the $O(a_s^4)$-level 
is  emphasised.  
\end{abstract}
 
\section{Introduction}
It is known that the results of perturbative calculations of the
physical quantities, which
obey the RG  equations (for the development of
the RG  method see e.g.  \cite{Stueckelberg:1951gg,
Bogolyubov:1956gh,Itzykson:1980rh}),
depend on the choice of the scale  and scheme of the  renormalization
procedure. 
In the case of QCD  this problem is of particular importance.
Indeed, calculations of the  multiloop   corrections  to  the
observable physical quantities and to  the related RG- functions
(namely,   $\beta$-function and various anomalous
dimensions) are usually performed  in the class  of minimal subtractions (MS)
schemes, and in the \MSbar-scheme \cite{Bardeen:1978yd}, in particular.
In this case, an  error of the comparison of theoretical results
with  experimental data
is usually determined
by varying the corresponding  renormalization scale $\mu^2$  within the
concrete interval, say $\mu^2/k \leq \mu^2 \leq k \mu^2$, where $k$ is the
conventionally
chosen number, i.e.  $k=2 \div 4$
(this convention was recently   used recently
in \cite{Alekhin:2012vu} ). 
As can be seen from this work and from the studies of heavy flavour
contributions to DIS sum rules \cite{Blumlein:1998sh}, 
this interval for  $k$ is indeed conventional. 
Say, 
the analysis of  \cite{Blumlein:1998sh} motivates
the choice of the following interval for $\mu^2$:
$m_q^2\leq \mu^2\leq (6.5m_q)^2$, where $m_q$ are  the $c$ and $b$-quark pole
masses. Note, that in the process  fitting the CCFR 
collaboration  $xF_3$ structure functions
data for  $\nu N$ DIS  \cite{Kataev:2001kk}  both
ways  of fixation  of scale variations were considered for estimating
a theoretical error-bar of the extracted expression for
$\alpha_s^{\MSbar}(M_Z)$. 
The first one was used in the case when the number of flavours was  fixed as  $n_f=4$,  
while the second one  was used to    estimate  the sensitivity 
of the fitted  results to their transformation from
$n_f=4$ numbers of flavours  to the energy region with
$n_f=5$ numbers of flavours.

 Taking into account higher order perturbative corrections
in the non-asymptotic QCD  regime
decreases, as a rule,  theoretical errors  of
the analysed quantities and the extracted QCD parameters, 
which arise from the variations of scales within the conventionally chosen
interval of values.    
In spite  of this, there is quite understandable  desire to formulate more concrete
theoretical prescriptions for  analysing scale-scheme
dependence  uncertainties  using the RG-based language. 
Among the  most applicable at present   methods are  
the Principle of Minimal Sensitivity (PMS) \cite{Stevenson:1981vj},  
the Effective Charges (ECH)   \cite{Grunberg:1982fw}
and the Brodsky-Lepage-McKenzie (BLM) approaches   \cite{Brodsky:1982gc}.
The first two of them are based on the concepts of scheme-invariant
quantities. Both ECH and PMS approaches are widely
 used in the concrete phenomenological studies
(see e.g. 
\cite{Chyla:1991ca,Chyla:1992cg,Maxwell:2011dy,Stevenson:2012ti,Posolda:2014vxa}). 
In the process of these studies the
gauge-invariant and vertex dependent  schemes of defining
the QCD coupling constant  are usually used.
These  schemes include  
the original  MS-scheme \cite{'tHooft:1973mm},
its  G-scheme  \cite{Chetyrkin:1980pr} and  the \MSbar-scheme
\cite{Bardeen:1978yd} variants, 
and the  number of other similar
MS-like schemes, which are unified
within the class of $R_{\delta}$-schemes \cite{Mojaza:2012mf}.
Since all of them are
related to the dimensional regularization \cite{'tHooft:1972fi} and are
gauge-invariant,
there are no problems with scale-scheme ambiguities of the BLM-approach,
discussed \cite{Celmaster:1982zj,Chyla:1995ej} in the
case of various momentum subtraction (MOM) schemes, which in QCD
depend on the gauge choice.

\section{The  generalizations of the BLM approach beyond the NLO:
polarized Bjorken sum rule as a  typical example}

The
first NNLO  generalization of the BLM approach was formulated in
\cite{Grunberg:1991ac}. It was shown  that it is possible
to absorb unambiguously  all $n_f$-dependent terms from the NNLO corrections to
the RG-invariant measurable quantities,
evaluated in the \MSbar-scheme,  by introducing
the  coupling dependence correction  $O(\alpha_s)$ to  the BLM 
scale $\mu^2_{BLM}$ 
fixed from the   NLO approximations of the considered physical
quantities.
Note  that this feature of the generalized BLM prescriptions was
confirmed in the
process of  incomplete all-order  extension
of the BLM approach \cite{Ball:1995wa}  aimed at  resummation of the
renormalon-type
terms $(\beta_0\alpha_s)^n$  to the BLM
scales of the perturbative series for the  $\tau$- hadronic width
and of the  relations between the  pole and running heavy quark masses.

Now let us now consider  modern approaches of  formulating  all-order
generalizations  of the BLM method  using
the N$^3$LO  approximation for the Bjorken sum rule of the
polarized lepton-nucleon scattering  as an example.
This sum rule  is defined as
\begin{equation}
S_{Bjp}=\int_0^1 g_1^{lp-ln}(x,Q^2)dx= \frac{g_A}{6}C^{Bjp}(a_s)   .
\end{equation}
The  function $C^{Bjp}$ contains the non-singlet (NS)  and singlet (SI)
contributions
$C^{Bjp}(a_s)=C^{\rm Bjp}_{\rm NS}(a_s)+C^{\rm Bjp}_{\rm SI}(a_s)$ .
The existence  of the SI  term at the $O(\alpha_s^4)$ level was
demonstrated in \cite{Larin:2013yba}. Its concrete  analytical
expression is not yet fixed  by  direct diagrammatic calculations. 
Since  we are  interested in applications
of the generalized BLM approaches of \cite{Kataev:2014jba}, which
are  similar to the seBLM method
\cite{Mikhailov:2004iq},  and of the PMC approach (see 
recent  reviews
\cite{Wu:2013ei,Wu:2014iba}),  we will
neglect this SI-type $a_s^4$- term and consider the  expression
for $C^{\rm Bjp}_{\rm NS}(a_s)$.

Both methods  start  with the application of the
the $\{\beta\}$-expansion  of the perturbative \MSbar-scheme
coefficients for  $C^{\rm Bjp}_{\rm NS}(a_s)$. Within the  seBLM-motivated 
approach of
\cite{Kataev:2014jba} this concrete structure of
the $\{\beta\}$-expansion was  obtained in \cite{Kataev:2010du}
from the \MSbar-scheme generalizations of the
original Crewther relation \cite{Crewther:1972kn} 
fixed at the NNLO in \cite{Broadhurst:1993ru} and at the N$^3$LO in
\cite{Baikov:2010je} by  using the  $\{\beta\}$-expanded $O(a_s^4)$
representation for the $e^+e^-$ characteristic, namely, the 
$C_D^{NS}(Q^2)$ function \cite{Mikhailov:2004iq}.

The expression for $C^{\rm Bjp}_{\rm NS}(a_s)$ at the N$^3$LO level
\begin{equation}
\label{CNS}
C^{\rm Bjp}_{\rm NS}(a_s)=1+\sum_{i}^{4}c_i a_s^{i}
\end{equation}
was calculated in \cite{Baikov:2010je} in the \MSbar scheme, $a_s=\alpha_s/\pi$.
Using the $\{\beta\}$-expansion formalism  
we express  the coefficients $c_i$ as \cite{Kataev:2010du}
 \begin{eqnarray}
\label{c1b}
c_1&=&c_1[0], \\
\label{c2b}
c_2&=&\beta_0(n_f)c_2[1]+c_2[0], \\ \nonumber
c_3&=&\beta_0^2(n_f)c_3[2]+\beta_1(n_f)c_3[0,1] \\ \label{c3b}
&&+ \underline{\beta_0(n_f)c_3[1]}+c_3[0], \\ \nonumber
c_4&=&\beta_0^3(n_f)c_4[3]+\beta_1(n_f)\beta_0(n_f)c_4[1,1 ] \\ \nonumber
&&+\beta_2(n_f)c_4[0,0,1]+
\beta_0^2c_4[2]  \\ \label{c4b}
&&+ \underline{\beta_1(n_f)c_4[0,1]}+
\underline{\beta_0(n_f)c_4[1]}+c_4[0],
\end{eqnarray}
while certain values of the elements in the RHS are fixed \cite{Kataev:2010du,Kataev:2014jba}
following to generalized Crewther relation.
Here $n_f$ is the number of
fermion  flavours and $\beta_i(n_f)$ 
are the coefficients of the QCD $\beta$-function in the MS-like schemes,
which are  defined as
\begin{equation}
\label{betaf}
\mu^2\frac{da_s}{d\mu^2}=\beta(a_s)=-\sum_{i\geq 0}\beta_{i}(n_f)a_s^{i+2}
\end{equation}
In the \MSbar-scheme with one scale $\mu^2=Q^2$ the analytical
expressions for the NNLO $\{\beta\}$-approximation for $C^{\rm Bjp}_{\rm NS}(a_s)$
contain
the  {\it underlined} terms in Eq.(\ref{c3b}),(\ref{c4b}).
They are {\bf absent} in a similar $\{\beta\}$-representation
of the \MSbar perturbative expression for $C^{\rm Bjp}_{\rm NS}(a_s)$
(see e.g. Eq.(160) review  \cite{Wu:2013ei}).
The reason for this is more technical than theoretical.
As was shown in \cite{Mikhailov:2004iq}, at the NNLO it is possible to get
the $\{\beta\}$-expansion for the $e^+e^-$ $C_D^{NS}(a_s)$-function
with the
concrete  coefficients of the $\{\beta\}$-expanded pattern. It was possible to 
fix these terms 
only using an additional to  $n_f$  degree of freedom, namely,  the number
$n_g$ of multiplets of massless gluino.  Its analytical contribution to 
the NNLO correction of the
$C_D^{NS}(a_s)$-function is known from the results of
\cite{Baikov:2010je}, while the additional
contributions of $n_{gl}$  to the $\beta_0(n_f)$ and $\beta_1(n_f)$-functions
of QCD with gluino  are known from \cite{Clavelli:1996pz}.
Since  $\beta_0(n_f,n_{gl})$, $\beta_1(n_f,n_{gl})$ are liner in both  $n_f$ and
$n_{gl}$, this allows us to separate the contributions of $\beta_1$ and
$\beta_0$ to the NNLO correction of the $D^{NS}(a_s)$-function and obtain
the $\{\beta\}$-expansion pattern of the $O(a_s^3)$ coefficient with
4 terms  similar to the ones  entering  in Eq.(\ref{c3b}).

Note that without this additional information it is impossible to extract
 the contributions of $\beta_1(n_f)$ and $\beta_0(n_f)$ to the NNLO corrections
of physical quantities from the ordinary $n_f$-expansion of these terms. 
Indeed, the NNLO correction  contains three 
terms. In the case of polarized $C^{\rm Bjp}_{\rm NS}(a_s)$
the $O(a_s^3)$ coefficient  has the following form:
\begin{equation}
c_3(n_f)=c_{3,0}+c_{3,1}n_f+c_{3,2}n_f^2~.
\end{equation}
To avoid rather delicate and complicated studies,
 the authors of \cite{Wu:2013ei,Wu:2014iba} prefer to neglect
in the NNLO corrections of their  initial \MSbar-expressions
the term, proportional to the  single power of  $\beta_0$. 
The price for that
is the corruption of the structure of the  generalized Crewther relations
in the \MSbar-scheme
(see \cite{Kataev:2014jba} and  \cite{Kataev:2014zha} for a 
more   detailed discussion of this subject). Note  that
the generalized Crewther relations result from the  fundamental
properties of the
conformal symmetry and   its violation by the conformal  anomaly in QCD
(for the discussions see \cite{Braun:2003rp}).

Unfortunately, the absence  of any  information
about the calculated effects of the manifestation of
additional degrees of freedom (like the contributions of $n_{gl}$-multiplet
of gluinos)
in  the  analytical expression of   the $O(a_s^4)$
correction to $C_D^{NS}$, evaluated in \cite{Baikov:2010je}
in the case of   $SU(N_c)$ colour gauge group,  does  not
allow one   to extract  analytical
expressions for the coefficients  $c_4[1,1]$, $c_4[0,0,1]$,
$c_4[2]$, $c_4[0,1]$, $c_4[1]$ and $c_4[0]$ using the
ideas proposed in \cite{Mikhailov:2004iq}
(apart from  its $C_F^4$ contribution to $c_4[0]$,
first determined in \cite{Kataev:2008sk} from  the application of the
original Crewther relation \cite{Crewther:1972kn}, and  the leading term of
the $\beta_0$-expansion with the coefficient
$c_4[3]$, which  is known from the calculations
of \cite{Broadhurst:1993ru}).
Like the 
 coefficients
$c_3[2]$, $c_3[0,1]$, $c_3[1]$ and $c_3[0]$
analytically defined in \cite{Kataev:2010du}, 
these  still unknown terms are
crucial for demonstrating    the numerical
difference between different
generalizations of the BLM approach, namely between
the    PMC approach,  discussed e.g.  in 
\cite{Wu:2013ei},\cite{Wu:2014iba},  and 
single-scale  $\{\beta\}$-expanded generalization of  the BLM approach,
studied  in \cite{Kataev:2014jba}.   In view of the
lack  of knowledge
of the $\{\beta\}$-expanded expression  of the  coefficient $c_4$  in
Eq.(\ref{c4b}),  the single-scale  analysis of the
 $O(a_s^3)$-approximation of Eq.(\ref{CNS}) was considered  only. 
Let us have a look at  it
from another point of view, namely,  transforming  it
to   the case of  multiple scales, as was proposed  
in \cite{Mojaza:2012mf}.

\section{The scale-dependence of the
BLM generalizations at the NNLO and beyond}
Consider first the $\{\beta\}$-expanded form
of the  $O(a_s^4)$  approximation for the Bjorken
polarized sum rule   defined in the \MSbar-scheme
by Eqs.(\ref{CNS})-(\ref{c3b}) and transform it to the
multiple-scale case using the solution of the RG-equation
 presented in
\cite{Mojaza:2012mf} and written down in the form
 $R_{\delta}$
relation, defined in  \cite{Mojaza:2012mf} as 
\begin{equation}
\label{transf}
a_s(Q^2)=a_s(Q^2_{\delta})+\sum_{n=1}^{\infty}\frac{1}{n!}\frac{d^{n}
a_s(Q^2)}{d\ln Q^2}^n\Big|_{Q^2=Q^2_{\delta}}(-\delta)^n
\end{equation}
where $\ln Q^2/Q^2_{\delta}=-\delta$. This transformation relation  leads
to the  $O(a_s^4)$ multiscale approximation of $C^{\rm Bjp}_{\rm NS}(a_s)$,
which can be obtained from the straightforward   RG-transformations
of the power-series for $C^{\rm Bjp}_{\rm NS}(a_s)$ ( see e.g.
\cite{Mojaza:2012mf}), namely,
\begin{eqnarray}
\label{Q2}
C^{\rm Bjp}_{\rm NS}(a_s)&=&1+c_1[0]a_s(Q_1^2)+\bigg[\beta_0c_2[1]+
c_2[0]+\beta_0c_1[0]\delta_1\bigg]a_s^2(Q^2_2) \\ \nonumber
&+&\bigg[\bigg(\beta_0^2c_3[2]+\beta_1c_3[0,1]+\underline{\beta_0c_3[1]}+c_3[0]\bigg)+ \bigg(\beta_0^2\delta_1^2
+\beta_1\delta_1\bigg)c_1[0] \\ \label{Q3}
&+&2\beta_0\bigg(\beta_0c_2[1]+c_2[0]\bigg)\delta_2\bigg]a_s^3(Q^2_3)
\\ \nonumber
&+&\bigg[\bigg(\beta_0^3c_4[3]+\beta_1\beta_0c_4[1,1 ] +\beta_2c_4[0,0,1]+
\beta_0^2c_4[2] + \underline{\beta_1c_4[0,1]}+
\underline{\beta_0c_4[1]}+c_4[0] \bigg) \\ \nonumber
&+&\bigg(\beta_0^3\delta_1^3+ \frac{5}{2}\beta_1\beta_0\delta_1^2+\beta_2\delta_1\bigg)c_1[0]+\bigg(3\beta_0^2\delta_2^2+2\beta_1\delta_2\bigg)\bigg(
\beta_0c_2[1]+c_2[0]\bigg) \\
\label{Q4}
&+&3\beta_0\bigg(\beta_0^2c_3[2]+\beta_1c_3[0,1]+ \underline{\beta_0c_3[1]}+c_3[0]\bigg)\delta_3\bigg]a_s^4(Q^2_4) +O(a_s^5)
\end{eqnarray}
Here the scales are  defined as  $Q^2_K=Q^2exp(\delta_k)$.
Note  that the $\{\beta\}$-dependent structures of the $\delta_k$
expressions of the terms in Eqs.(\ref{Q2}), (\ref{Q3})
and Eq.(\ref{Q4}) ${\bf coincide}$ with the introduced in
\cite{Mikhailov:2004iq} $\{\beta\}$-expanded structure of
Eqs.(\ref{c2b})-(\ref{c4b}) introduced in \cite{Mikhailov:2004iq}.
This feature was
already observed in \cite{Mojaza:2012mf} and  \cite{Wu:2013ei}.
This fact is not 
{\bf accidental}  at all,  but follows from the general principles
of the   RG  method.
Moreover, contrary to the claims of refs.\cite{Mojaza:2012mf},
it is impossible to neglect in this expansion 
the terms  proportional to $\beta_0a_s^3(Q^2)$ in
Eq.(\ref{c3b}) (namely, to put {\bf  to zero}
the $c_3[1]$ coefficient in Eq.(\ref{c3b})
of the $\{\beta\}$-expanded expression for $C^{\rm Bjp}_{\rm NS}(a_s)$.
Indeed, this will automatically lead to disappearance of the
$3\beta_0^2r_{4,2}$ in Eq.(6) of \cite{Mojaza:2012mf}, which corresponds to
the 
$3\beta_0^2c_3[1]\delta_3a_s^4(Q_4^2)$-term,  underlined in part in  
 Eq.(\ref{Q4}).

To conclude, within the multiple-scale considerations of
\cite{Mojaza:2012mf},\cite{Wu:2013ei}, \cite{Wu:2014iba} the
neglected and   underlined in Eqs.(\ref{Q3}),(\ref{Q4}) terms will affect
the results for the scales $Q_3^2$ and $Q_4^2$, 
 obtained and discussed in \cite{Mojaza:2012mf},\cite{Wu:2013ei},
\cite{Wu:2014iba}.  They  should
be corrected by taking into account the  $\{\beta_i\}$-dependent terms,
underlined in
Eq.(\ref{Q3}),(\ref{Q4}).

\section{The prediction of the $\{\beta\}$-dependent NNLO
expression for the Bjorken polarized sum rule with
$n_{gl}$ multiplet of gluinos}

One of the ways to confirm the
$\{\beta\}$-dependent structure of the NNLO expression for $C^{\rm Bjp}_{\rm NS}(a_s)$
from Eqs.(\ref{c1b})-(\ref{c3b}), which is in agreement with the analytical
result of the direct  NNLO QCD  calculation of \cite{Larin:1991tj},
is to
evaluate analytically at the $a_s^3$-level   additional contributions from
the  $n_{gl}$ multiplets of SUSY QCD  gluinos, as   was done 
\cite{Chetyrkin:1996ez} in the case of the  $e^+e^-$ characteristic
$C_D^{NS}(Q^2)$. After this, it  will be extremely interesting  to use the ideas
of \cite{Mikhailov:2004iq} and  combine this possible new result
with the analytical expressions for   first two
coefficients of the RG $\beta$-function $\beta_0(n_f,n_{gl})$ and
$\beta_1(n_f,n_{gl})$. We believe, this possible
study will coincide with the  prediction  made in \cite{Kataev:2014jba} 
namely, with the following $O(a_s^3)$ approximation  for $C^{\rm Bjp}_{\rm NS}(a_s)$:
\begin{eqnarray}
\nonumber
C^{\rm Bjp}_{\rm NS}(a_s)&=&1-\frac{3}{4}C_Fa_s +\bigg(-\frac{3}{2}C_F\beta_0(n_f,n_{gl})
+\frac{21}{32}C_F^2-\frac{1}{16}C_FC_A\bigg)a_s^2 \\ \nonumber
&+&\bigg[-\beta_0^2(n_f,n_{gl})\frac{115}{24}C_F-\beta_1(n_f,n_{gl})\left(\frac{59}{16}
+3\zeta_3\right)C_F  \\ \nonumber
&+&\beta_0(n_f,n_{gl})\bigg(
\left(\frac{83}{24}-\zeta_3\right)C_F^2+\left(\frac{215}{192}-6\zeta_3+
\frac{5}{2}\zeta_5\right)C_F C_A\bigg) \\ \label{as3}
&+&-\frac{3}{128}C_F^3-\frac{65}{64}C_F^2C_A-\bigg(\frac{523}{768}-\frac{27}{8}
\zeta(3)\bigg)C_FC_A^2 \bigg] a_s^3 +O(a_s^4)
\end{eqnarray}
 where
\begin{eqnarray}
\label{beta0}
\beta_0(n_f,n_{gl})&=&\frac{11}{12}C_A-\frac{1}{3}\bigg(T_FN_F+
\frac{1}{2}n_{gl}C_A\bigg) \\ \label{beta1}
\beta_1(n_f,n_{gl})&=&\frac{17}{24}C_A^2-\frac{5}{12}C_A\bigg(T_FN_F+
\frac{1}{2}n_{gl}C_A\bigg)-\frac{1}{4}\bigg(T_FN_FC_F+\frac{1}{2}n_{gl}C_A^2\bigg)
\end{eqnarray}
are the coefficients of the corresponding $\beta$-function, defined by
Eq.(\ref{betaf}). The prediction of Eq.(\ref{betaf})
was obtained in  \cite{Kataev:2014jba} using a
similar expression for $C_D^{NS}(Q^2)$, which result  from the
calculations of \cite{Chetyrkin:1996ez}, detailed considerations
of \cite{Mikhailov:2004iq} and following from the   conformal symmetry 
original
Crewther relation  \cite{Crewther:1972kn}
between the  $C_D^{NS}(Q^2)$ and  $C^{\rm Bjp}_{\rm NS}(Q^2)$ coefficient 
functions. It was also checked in
\cite{Kataev:2014jba} that the same expression can be obtained from the 
 $O(a_s^3)$- generalization of
the Crewther relation,
discovered in \cite{Broadhurst:1993ru}. In this relation the 
defined by the conformal symmetry term  
is modified by the 
to the conformal symmetry
breaking term , namely by the factorized 
$\beta$-function of QCD with $n_{gl}$
multiplets of massless gluino. We are sure 
that a  direct check of the
prediction of Eq.(\ref{as3}) may give an  additional argument in favour of the
used in our works form of  $\{\beta\}$-expansion at the level of
NNLO corrections, considered in this report. Moreover, 
possible N$^3$LO evaluation of $C^{NS}(a_s)$ in QCD with $n_{gl}$  gluino
multiplet should clarify how to extract still unknown 
analytical coefficients of the $\{\beta\}$-dependence 
pattern of the $O(a_s^4)$ -correction to $C^{NS}(a_s)$.  

\section{Theoretical advantages of one-scale  seBLM/PMC and  its
phenomenological troubles}
Let us  transform the general multiple-scale
NNLO approximation for $C^{\rm Bjp}_{\rm NS}(a_s)$, defined in Eq.(\ref{Q2}),(\ref{Q3}),
to the single-scale approximation, studied in
\cite{Kataev:2014jba}. This can be simply done by fixing
$\delta_1$=$\delta_2$=$\delta$. Absorbing now the $\beta_0$- dependent term
into the scale $\delta_1$ at the $O(a_s^2)$ level we obtain the standard
BLM expression for $C^{\rm Bjp}_{\rm NS}$, namely
\begin{equation}
C^{\rm Bjp}_{\rm NS}(a_s)=1+c_1[0]a_s(Q_{BLM}^2)+c_2[0]a_s^2(Q_{BLM}^2)+O(a_s^3)~~~.
\end{equation}
Taking into account that   $c_1[0]=-\frac{3}{4}C_F=-1$  and
$c_2[1]=-\frac{3}{2}C_F=-2$ we get the value of the standard BLM scale
$Q^2_{BLM}=Q^2\rm{exp(-c_2[1]/c_1[0])}=Q^2exp[-2]=Q^20.135$.
The  $c_2[0]$ term 
was obtained in \cite{Kataev:2014jba}. Its expression is  
$c_2[0]=-
\frac{21}{32}C_F^2+\frac{1}{16}C_FC_A=-11/12=-0.91(6)$.

In the work of Ref.\cite{Kataev:2014jba}
 we absorb into the BLM scale  of the NNLO   $O(a_s^3(Q_{BLM}^2)$
single-scale coefficient from  Eq.(\ref{Q3})  
the terms  proportional to $\beta_0^2c_3[2]$, $\beta_0c_3[0,1]$, 
$\beta_0c_3[1]$ 
as well. To avoid lengthy
discussions, given in the original work  \cite{Kataev:2014jba}
we will consider the case of $N_F=3$ number of massless flavours and
present only  part of
analytical and numerical expressions of the corresponding
coefficients of the $\{\beta\}$-expansion procedure,
which  define  the BLM scale of the NNLO approximation
for $C^{\rm Bjp}_{\rm NS}(a_s)$.  They are taken from the  results of
\cite{Kataev:2014jba} and  read
$c_3[2]=-\frac{151}{24}C_F= -115/18=-6.38(8)$,
$c_3[0,1]=(-\frac{59}{16}+3\zeta_3)C_F=
(-\frac{59}{12}+4\zeta_3)=-0.108$ and $c_3[1]=(\frac{83}{24}-\zeta_3)C_F^2+
(\frac{215}{192}-6\zeta_3+\frac{5}{2}\zeta_5)C_FC_A=\frac{4591}{432}-
\frac{232}{9}\zeta_3+10\zeta_5=-9.989$. Using these numbers and 
the expressions for $c_1[0]$, $c_2[0]$ and $c_2[1]$,  as given above,
we obtain    the expression  for  the NNLO BLM
 scale for  $C^{\rm Bjp}_{\rm NS}(a_s)$, given in \cite{Kataev:2014jba} in a bit
different normalization
\begin{equation}
\label{NNLO}
Q^2_{NNLO}=Q_{BLM}^2exp[-7.32\beta_0a_s(Q_{BLM}^2)]=
Q^2exp[-2-7.32\beta_0a_s(Q_{BLM}^2)]~~~.
\end{equation}.  
The
expressions for the $O(a_s^3(Q^2_{NNLO}))$ coefficient to
$C^{\rm Bjp}_{\rm NS}(a_s)$ read \cite{Kataev:2010du}:
\begin{equation}
c_3[0]=-\frac{3}{128}C_F^3-\frac{65}{64}C_F^2C_A-\bigg(\frac{523}{768}-
\frac{27}{8}\zeta_3\bigg)C_FC_A^2 \approx 35.034
\end{equation}
where the numerical value is given in the case of $SU(N_c=3)$
colour gauge group. Thus, the final result we are interested in reads
\begin{equation}
\label{CNSBLM}
C^{\rm Bjp}_{\rm NS}(a_s)=1-a_s(Q_{NNLO}^2)-0.91(6)a_s^2(Q_{NNLO}^2)+35.034a_s^3(Q_{NNLO}^2)+O(a_s^4)
\end{equation}
Considering the one-scale  PMC-type expression we conclude that
\begin{itemize}
\item
its analytical coefficients extracted from the $\{\beta\}$-expansion
procedure  of  \cite{Mikhailov:2004iq} and the \MSbar-scheme
generalizations of the Crewther relation of \cite{Broadhurst:1993ru},
\cite{Kataev:2010du} do not depend on  the number of flavours and on  the
terms  proportional to the QCD $\{\beta\}$-function. Combined with 
similar expressions for the coefficients $d_i[0]$ of the
$\{\beta\}$-  expansion representation of the \MSbar-scheme expression
for the $e^+e^-$ characteristic $C_D^{NS}(a_s)$ these coefficients
satisfy the scheme-independent relations, which follow from the 
original Crewther relation of \cite{Crewther:1972kn}, which 
is  based on the
conformal symmetry.
\item
Unfortunately, in view of not small $O(a_s^3)$-term
 application of the single-scale realisation of the PMC
approach of Eq.(\ref{CNSBLM}),
 which is very similar to the seBLM method, spoils the satisfactory
convergence of the $O(a_s^3)$ \MSbar-scheme approximation
of the polarized  Bjorken sum rule perturbative QCD expression, while
the NNLO generalization of the  BLM-scale of Eq.(\ref{NNLO}) moves
the applicability of the resulting perturbative expressions to the
region of higher energies.
\end{itemize}

\section{Conclusions}
At the current stage of  our studies, presented  in detail in
\cite{Kataev:2014jba}, we discover several phenomenological disadvantages
of the applicability of the  BLM approach,  
generalized to $O(a_s^3)$-level even following the 
application of the   $\{\beta\}$-expansion formalism, proposed in
\cite{Mikhailov:2004iq} and studied in \cite{Kataev:2010du}. 
We raise
the question of the inappropriate use of this formalism within  
the Principle of Maximal Conformality 
considered in \cite{Mojaza:2012mf,Wu:2013ei,Wu:2014iba} and propose the calculating test,
which may give extra argument in favour of the self-consistency
of the structure of the $\{\beta \}$-expansion approach, as used by us at the
$O(a_s^3)$-level. We would like also to mention that
it is possible to invent the way  how to make the generalized BLM
approach more suitable for high-energy studies. However,  
we are still  unable to give any straightforward theoretically
motivated prescription for its formulation.  At present,
this  work, which was started in \cite{Kataev:2014jba}, is based on some
empirical observations and necessitates more careful analysis of
$O(a_s^4)$-contributions to the RG-invariant quantities,
still unknown within the used version of the theoretically consistent
$\{\beta\}$-expansion approach. Note, that our studies are also of importance 
in view  of the necessity of better understanding of the applications of the 
NNLO generalizations of the BLM approach to other imprtant observables, 
like the event-shape distributions \cite{Gehrmann:2014uva},   which are 
 measured precisely at $e^+e^-$-colliders. 
.

\section{Acknowledgements}
I would like to thank  Sergey Mikhailov for our  
productive collaboration and for useful  discussions of the 
material of this talk.
This work, as well as the participation at the ACAT-20014 Workshop 
was supported in part by RSCF, Grant N 14-22-00161. I am  
also grateful to the organizers of    this  Workshop and in
particular to  Milos Lokajicek for the warm   
hospitality in Prague and partial
financial support.

\section*{References}


\begin{thebibliography}{3}
\bibitem{Stueckelberg:1951gg}
Stueckelberg E C G and Petermann A 1951 {\it Helv. Phys. Acta} {\bf 24} 317
\bibitem{Bogolyubov:1956gh}
 Bogolyubov N N and  and Shirkov D V 1956 {\it Nuov. Cim}. {\bf
3} 845
\bibitem{Itzykson:1980rh} Itzykson C and J. B. Zuber J B
{\it Quantum Field Theory} 1980  (New York,  McGRAW-HILL)
\bibitem{Bardeen:1978yd}
Bardeen W A, Buras A J, Duke W and Muta T
1978  {\it Phys. Rev.} D {\bf 18} 3998
\bibitem{Alekhin:2012vu}
Alekhin S, Blumlein J, Daum K,  Lipka K  and Moch S 2013
{\it Phys. Lett.} B {\bf 720} 172 (Preprint hep-ph/1212.2355)
\bibitem{Blumlein:1998sh}
Blumlein J  and van Neerven W L 1999   {\it Phys. Lett.}  B {\bf 450} 417
(Preprint hep-ph/9811351)
\bibitem{Kataev:2001kk}
 Kataev A L, Parente G  and Sidorov A V 2003
{\it   Phys. Part.Nucl.}   {\bf 34}  20 (Preprint hep-ph/0106221)
\bibitem{Stevenson:1981vj}
Stevenson P M 1981  {\it Phys. Rev.} D  {\bf 23}  2916
\bibitem{Grunberg:1982fw}
Grunberg G 1984  {\it Phys. Rev.} D {\bf 29} 2315 (Preprint
Ecole Polytechnique 82-0721)
\bibitem{Brodsky:1982gc}
Brodsky S J , Lepage  G P and Mackenzie B P
1983 {\it Phys.Rev.}  D {\bf 28}  228.
\bibitem{Chyla:1991ca}
 Chyla J,  Kataev A  and Larin S A  1991
{\it Phys.Lett.} B {\bf 267}  269 (Preprint UM-TH-91-06, DESY 91-059)
\bibitem{Chyla:1992cg}
 Chyla J  and Kataev A L 1992
{\it Phys. Lett.}  B {\bf 297}  385 (Preprint CERN-TH-6604-92, hep-ph/9209213)
\bibitem{Maxwell:2011dy}
 Maxwell C J   and Morgan K  E 2012
{\it   Nucl.  Phys.}  B {\bf 858} 405 (Preprint hep-ph/1108.6204)
\bibitem{Stevenson:2012ti}
Stevenson P M  2013
{\it Nucl.Phys.}    B {\bf 868}  38 (Preprint hep-ph/1210.7001)
\bibitem{Posolda:2014vxa}
  P.~Posolda, {\it J.\ Phys.}  G {\bf 41} (2014) 095007.
\bibitem{'tHooft:1973mm}
't Hooft G 1973 {\it Nucl. Phys.} B {\bf 61} 455
\bibitem{Chetyrkin:1980pr}
Chetyrkin K G, Kataev A L  and Tkachov F V
1980 {\it Nucl. Phys.}  B {\bf 174} 345
\bibitem{Mojaza:2012mf}
 Mojaza M,  Brodsky S J  and Wu X G 2013
{\it Phys. Rev. Lett.}  {\bf 110} 192001 (Preprint hep-ph/1212.0049)
\bibitem{'tHooft:1972fi}
 't Hooft G  and Veltman  M J G  1972
{\it Nucl. Phys.} B {\bf 44} 189.
\bibitem{Celmaster:1982zj}
Celmaster W  and Stevenson P M  1983
{\it Phys. Lett.}  B {\bf 125}  493.
\bibitem{Chyla:1995ej}
 Chyla J  1995 {\it Phys. Lett.} B {\bf 356} 341 (Preprint hep-ph/9505408).
\bibitem{Grunberg:1991ac}
Grunberg G  and Kataev A L 1992
{\it   Phys. Lett.}   {\bf  279} 352.
\bibitem{Ball:1995wa}
Ball P, Beneke M  and Braun V M 1995
{\it   Phys. Rev.} D {\bf 52}  3929 (Preprint  hep-ph/9503492)
\bibitem{Larin:2013yba}
 Larin S A {\it Phys. Lett.} 2013 B {\bf 723}  348
\bibitem{Kataev:2014jba}
Kataev A L  and Mikhailov S V 2015    Generalization o0f the Brodsky-Lepage-Mackenzie optimization within 
the $\{\beta\}$-expansion and the Principle of Maximal Conformality,
{\it Phys. Rev.} D  {\bf 91} 014007;  
(Preprint INR-TH-9-2014; hep-ph/1408.0122)
\bibitem{Mikhailov:2004iq}
Mikhailov S V 2007 Generalization of BLM procedure and its
scales in any order of pQCD: a practical approach  J High Energy
Physics
{\it JHEP} {\bf 0706} 009 (Preprint JINR-E2-2005-25, hep-ph/0411397)
\bibitem{Wu:2013ei}
 Wu X G, Brodsky S J  and Mojaza M 2013
{\it   Prog. Part. Nucl. Phys.}   {\bf 72}  44 (Preprint hep-ph/1302.0599)
\bibitem{Wu:2014iba}
Wu X G,   Ma Y,   Wang S Q,   Fu H B,   Ma Y Y , Brodsky S J  and
Mojaza M 2014 Renormalization group invariance and optimal renormalization
scale-setting  Preprint SCAL-PUB-1593, hep-ph/1405.3196
\bibitem{Kataev:2010du}
 Kataev A L  and Mikhailov S V  2012
{\it Theor. Math. Phys.}  {\bf 170}  139 (Preprint hep-ph/1011.5248)
\bibitem{Crewther:1972kn}
 Crewther R J 1972
 {\it Phys. Rev. Lett.}    {\bf 28}  1421.
\bibitem{Broadhurst:1993ru}
Broadhurst D J  and Kataev A L  1993
{\it  Phys. Lett.}   B {\bf 315}  179 (Preprint hep-ph/9308274)
\bibitem{Baikov:2010je}
 Baikov P A, Chetyrkin K G    and Kuhn J H  2010
{\it Phys. Rev. Lett.}  {\bf 104}  132004 (Preprint hep-ph/1001.3606)
\bibitem{Chetyrkin:1996ez}
 Chetyrkin K G 1997
{\it Phys.Lett.} B  {\bf 391}  402 (Preprint hep-ph/9608480)
\bibitem{Clavelli:1996pz}
 Clavelli L,  Coulter P W  and Surguladze L R 1997
{\it Phys. Rev.Lett.} D {\bf 55} 4268 (Preprint hep-ph/9611355)
\bibitem{Kataev:2014zha}
Kataev A L  and Mikhailov S V 2014 $\{\beta\}$-expansion in QCD, its
conformal  symmetry limit: theory+applications, Preprint
hep-ph/1410.0554
\bibitem{Braun:2003rp}
 Braun V M, Korchemsky G  P  and Mueller D 2003
{\it Prog. Part. Nucl. Phys.}  {\bf 51}  311 (Preprint hep-ph/0306057)
\bibitem{Kataev:2008sk}
 Kataev A L 2008
{\it Phys. Lett.} B {\bf 668}  350 (Preprint hep-ph/0808.3121)
\bibitem{Larin:1991tj}
 Larin S A  and Vermaseren J A M  1991
{\it  Phys. Lett.} B {\bf 259}  345.
\bibitem{Gehrmann:2014uva}
Gehrmann T , Häfliger N and Monni P F 2014
{\it Eur. Phys.J.}  C {\bf 74} 2896
(Preprint   hep/ph:1401.6809)

\end{thebibliography}
\end{document}